\documentclass[sigconf]{acmart}
\usepackage{color, colortbl}

\AtBeginDocument{%
  \providecommand\BibTeX{{%
    \normalfont B\kern-0.5em{\scshape i\kern-0.25em b}\kern-0.8em\TeX}}}

\acmYear{2022}\copyrightyear{2022}
\setcopyright{rightsretained}
\acmConference[DBTest '22]{9th International Workshop of Testing Database Systems}{June 17, 2022}{Philadelphia, PA, USA}
\acmBooktitle{9th International Workshop of Testing Database Systems (DBTest '22), June 17, 2022, Philadelphia, PA, USA}
\acmPrice{}
\acmDOI{10.1145/3531348.3532177}
\acmISBN{978-1-4503-9353-9/22/06}

\begin{document}

\title{Journey of Migrating Millions of Queries on The Cloud}

\author{Taro L. Saito, Naoki Takezoe, Yukihiro Okada, Takako Shimamoto\\
Dongmin Yu, Suprith Chandrashekharachar, Kai Sasaki, Shohei Okumiya\\
Yan Wang, Takashi Kurihara, Ryu Kobayashi, Keisuke Suzuki\\
Zhenghong Yang, Makoto Onizuka}
\authornote{Osaka University, Osaka Japan}
\affiliation{%
  \institution{Treasure Data}
  \city{Mountain View}
  \state{CA}
  \country{USA}\\
}

\renewcommand{\shortauthors}{Taro and Naoki, et al.}

\begin{abstract}

Treasure Data is processing millions of distributed SQL queries every day on the cloud. Upgrading the query engine service at this scale is challenging because we need to migrate all of the production queries of the customers to a new version while preserving the correctness and performance of the data processing pipelines. To ensure the quality of the query engines, we utilize our query logs to build customer-specific benchmarks and replay these queries with real customer data in a secure pre-production environment. To simulate millions of queries, we need effective minimization of test query sets and better reporting of the simulation results to proactively find incompatible changes and performance regression of the new version. This paper describes the overall design of our system and shares various challenges in maintaining the quality of the query engine service on the cloud.

\end{abstract}






\maketitle

\section{Introduction}

Treasure Data is an enterprise customer data platform (CDP) on the cloud. Our customers continuously collect data from various sources into the CDP and analyze the activities of their own customers through the collected logs. The imported data is stored in a columnar storage so that the customers can make efficient distributed SQL queries. The analyzed data is used for building user segments that have some common characteristics (e.g., people visited a product web site and purchased the product before, or belonging to some age group, living area, etc.). Marketers, who are responsible for sending product advertisement to customers, need to find good segments of customers for optimizing their marketing campaigns so that they can target right users, who are most likely to purchase additional products and lead to the revenue growth.

As of February 2022, Treasure Data is processing more than 1.5 million SQL queries every day coming from 5,000+ users in various regions, including US, EU, Japan, Korea, etc. The number of processed records exceeds 100 trillion rows/day, which is equivalent to processing 1.2 billion rows/sec. Our job as a service provider is preserving the behavior of these SQL queries while maintaining the underlying query engine versions up-to-date so that our customers can keep processing their data analysis pipelines without worrying about maintaining the data platform.

\begin{figure*}[h]
\centering
\includegraphics[width=.80\textwidth]{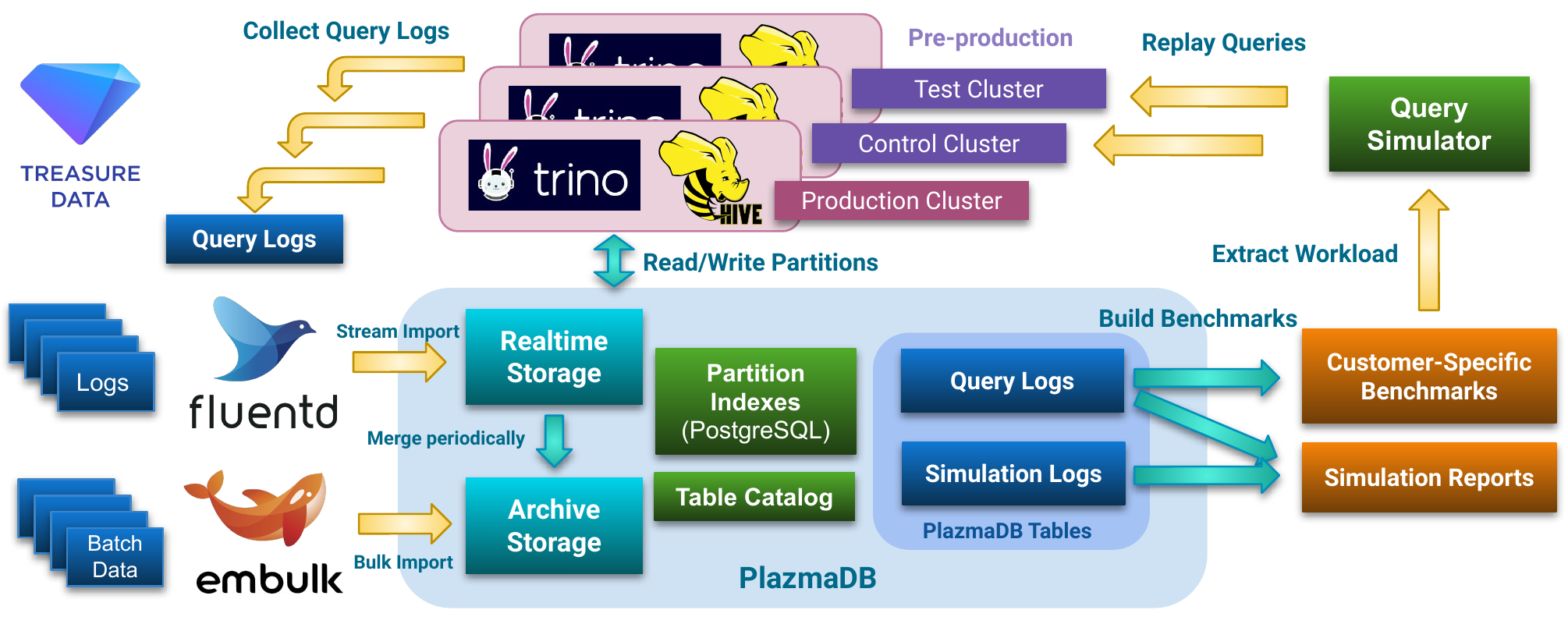}
\caption{The architecture of Treasure Data, including PlazmaDB, query engines, and query simulator}
\label{fig:overview}
\end{figure*}

\subsubsection*{Maintaining A Data Platform on the Cloud}


Treasure Data provides query engine services on the cloud by integrating open-source (OSS) distributed SQL query engines Trino (formerly known as Presto) \cite{Trino} and Hive \cite{10.1145/3299869.3314045} on top of Amazon Web Service (AWS). Our target customers (e.g., marketers) are not dedicated system engineers, so they have a strong demand to have a data platform which is easy to use and has no need to maintain by themselves. 
Treasure Data provides a web console and APIs for running SQL queries so that the users can analyze their data without caring too much about maintaining the data platform on the cloud. Our customers build complex data processing pipelines on top of our platform and depend on the stability of the query engine service to continuously extract insights from the data.

When using OSS query engines like Trino, we need to be prepared for the frequent updates of the query engine versions. Since its first release in 2013, Trino has already released 370 versions until February 2022, having a new release almost every week \cite{Trino}. Unlike proprietary software, these OSS products are developed by community, so our engineers are not always be able to review all of the relevant code changes. When OSS projects are actively developed like Trino, service providers like us need to carefully choose a version to use because a new version might include incompatible changes that might affect the behavior of our customer queries and the performance.




To verify the correctness and performance stability of customer queries in new query engine versions, we developed a query simulator, which can replay customer queries using real customer data sets. The query simulator complements other quality control methods in our release pipeline, such as unit tests, integration tests, and end-to-end (e2e) tests \cite{10.5555/3288797}. Our query simulator runs customer-specific SQL benchmarks built from the historical query logs. This ensures the correctness of the customer query results and the performance stability of individual queries before shipping a new version of the software to production. This approach enabled us to choose the most stable version for our customers.


For simulating millions of queries processed per day, first, we need to continuously collect large-scale query logs even if the format of query logs changes over time. It is also important to extract typical workloads of individual customers to reduce the number of queries to test. Our approach for these problems is using our own service for collecting and analyzing query logs. An analysis of our historical query logs revealed that more than 97\% of queries of our customers are recurrent ones, running hourly, daily, weekly, monthly, etc. By clustering such recurrent queries, we can significantly reduce the query set size fed to our query simulator. In addition, to protect the privacy of the customer data, the query simulator obfuscates the query results by embedding a checksum computation process and ensures the containment of all intermediate data within a pre-production environment, which is isolated from the production customer traffic.

Simulating customer-specific query patterns has enabled us to fill the gap of responsibilities between service providers like Treasure Data, who need to manage millions of queries every day, and the customers, whose business depends on the result of a single-specific query. In the early days, we used only pre-defined query sets for testing query engines. As the usage grew, we started seeing errors that happened only at production queries after releasing new versions. To make the query testing more proactive, we introduced the customer-specific query simulation not only for detecting bugs before the release, but also for reporting individual queries that will be affected by upcoming version upgrades. Especially, when we need to introduce incompatible behaviors, such as SQL syntax changes, feature deprecation, performance regression due to optimization rule changes, etc., we need to notify the customers in advance so that they can have sufficient time to fix their queries. 
Explicitly reporting specific queries whose behavior will change in an upcoming version is more informative than providing release notes that simply describe all of the functionality changes. 

\subsection{Related Work}

Testing DBMSs using standard benchmarks TPC-H, TPC-DS has been a common practice. 
Trino OSS query engine provides trino-verifier utility \cite{Trino} for comparing such benchmark query results between multiple versions, but it doesn't support result obfuscation. Tableau reported a more extensive approach using a gold-standard benchmark consisting of 60k public workbooks to test their HyPer in-memory DBMS \cite{hyper}. Such static benchmarks, however, are not always a representative workload of customers. Recent approaches for testing customer-specific workloads include Snowtrail for testing Snowflake SQL queries \cite{10.1145/3209950.3209958} and DIAMetrics \cite{10.14778/3415478.3415551}, which also covers non-SQL workloads in Google. A formal framework for constructing a representative workload from query logs is studied in \cite{10.14778/3430915.3430931}, which requires the expert knowledge to select features to be tested. In this paper, we argue the importance of collecting query logs even for testing emerging OSS query engines using unsupervised methods. In this regard, we shed lights to the system for collecting query logs and practical applications of these query logs for monitoring and testing query engines. Morpheus \cite{199313} reported a method for defining SLOs of query performance induced from large query graphs. In this paper, we present a simple approach for observing customer expectations as implicit SLOs, which can be used if query logs are available in queryable storage. 

\smallskip

In the rest of this paper, we present the architecture of our service, including log collection system and query simulator (Section \ref{sec:architecture}). Section \ref{sec:analyzing} presents methods for utilizing query logs for building customer-specific benchmark. Section \ref{sec:migrating} navigates readers to our journey of migrating millions of queries to update the query engine versions. Finally, we discuss various challenges in operating query engine services on the cloud (Section \ref{sec:discussion}) and conclude the paper (Section \ref{sec:conclusions}). Although the code of our service and query simulator is proprietary, we believe that the methods and our best practices described in this paper are still useful for practitioners who want to build their own data platform on the cloud.


\section{System Architecture}
\label{sec:architecture}

Figure \ref{fig:overview} illustrates an overview of our system. In this section, we will see the design of our systems for collecting and querying data with distributed SQL query engines. We use our own service for collecting query logs and utilize these logs for building customer-specific SQL benchmarks and monitoring query performance (Section \ref{sec:analyzing}).

\subsection{PlazmaDB: Transactional Storage}

The core storage of Treasure Data is PlazmaDB\footnote{For the uniqueness, it is intentionally named as Plazma, not Plasma}, which is a time-series DBMS for collecting data from stream and batch data sources. To feed data into PlazmaDB, we have developed various open-source data connectors, such as Fluentd \cite{fluentd} for stream data ingestion and Embulk \cite{embulk} for bulk data ingestion. Both Fluentd and Embulk are extensible with plug-ins, so we can import various data from third-party services to Treasure Data. Tables in PlazmaDB always have a \emph{time} column for describing the event time in Unix time (epoch seconds since 1970-01-01). When the time column is missing in the input data, the ingestion time will be used. As most of the users usually analyze recently ingested data, they can accelerate their queries by specifying recent time ranges.

PlazmaDB has two types of storage, called real-time and archive storage. Stream ingested data is stored first in the real-time storage. Files in the real-time storage are usually small and fragmented as their sources are mobile devices, web access logs, etc. To make the query processing efficient, we periodically merge records in the real-time storage to create a more efficient size of columnar partitions, and move them to the archive storage. Typically a single partition file in the archive storage contains less than 1 million rows or 256MB in size after columnar compression. 

We use AWS S3 key-value storage for storing columnar partition files. Our query engines issue billions of S3 GET requests per day. Although S3 is fast enough for looking up individual objects, listing partition files in a specific path is significantly slow, so we manage additional indexes to these partition files using a PostgreSQL database. These extra indexes enable listing up millions of partitions on S3 in a few seconds, and it also supports transactional update of partition entries, which is used for implementing atomic SQL operations, such as INSERT, DELETE, etc. For example, while INSERT INTO queries are running, no other users see non-committed partitions. 

PlazmaDB uses MessagePack \cite{msgpack} as the underlying data format. MessagePack is a self-describing binary data format, in which every value has a type prefix (e.g., integer, float, boolean, string, array, map, etc.). For partition files stored on S3, we use a columnar version of MessagePack, called MPC1, whose column blocks have compressed sequences of MessagePack data. MPC1 format is similar to the existing columnar open-data formats such as Parquet \cite{parquet}, ORC \cite{orc}. One of the reasons we are using our own data format is that both Parquet and ORC didn't exist when we started the service in 2011, and also Parquet/ORC are schema-full data format, which has no support for schema-on-read conversions, which will be discussed below.





\subsection{Collecting Evolving Query Logs}

We use PlazmaDB not only for collecting data from our customers, but also for collecting internal metrics and logs of our services. PlazmaDB has been collecting trillion of query log records since when we started our service in 2011, and it already has 10 years of query logs. These query logs are valuable resources for understanding the usage of our customers and optimizing the query engine services. 

A unique aspect of PlazmaDB that enabled collecting query logs for such long years is its support for schema evolution and escalation; Tables in PlazmaDB require no predefined schema because the schema grows automatically as the user ingests new types of records (schema evolution). In addition, even if the data type of a column changes over time (schema escalation), we can safely embed different types of data values in the same column using self-describing MessagePack format. As we can keep collecting data without managing table catalogs, PlazmaDB has been used widely across Treasure Data to collect various internal service logs and metrics.

Allowing schema evolution and escalation is our design choice so as not to block the data ingestion. This is because we can add a data cleansing process after the data is ingested to the system, but we can't recover records if the data is rejected before reaching the system. This approach has been working well; historically, our query log tables have added many columns for new query parameters, especially when we had major query engine upgrades. We also have observed cases where MessagePack-based data format helped our customer data, for example, when some integer column needs to accept string values as well, or when unexpected data happen to be included in a column. Unlike traditional DBMSs, which require strict schema validation when inserting new records, PlazmaDB is flexible enough for the change of the characteristics of the data over time.\footnote{If necessary, the user can explicitly set specific column types to restrict unexpected schema escalations.} 

To read columns that might have schema escalation, our query engines implement schema-on-read data processing for converting physical data types in data files into logical data types specified in a table schema or SQL queries. For example, values 100 (integer) and "100" (string) have different physical representations, but if a query requires integer values, the system translates the string value "100" into an integer value 100. By looking at type prefixes of MessagePack data, we can select appropriate schema-on-read converters at ease. If such conversion is not possible (e.g., non-integer strings like "abc"), the returned value will be just null.\footnote{Even if invalid column values are present in physical partition files, recent query engines like Trino have better support of null value handling, so in our experiences, it will not distract analytical SQL processing.}






\subsection{Query Engines}

To provide SQL-based query engine services, we extended the OSS versions of Trino and Hive to access partition files stored in PlazmaDB with schema-on-read support. In addition, we support multi-tenant configurations to share the same cluster between multiple users, strict access control with database/table/column-level permissions, controlling query performance parameters based on the pricing plan of the users, etc. To collect query logs, we extended the internal event handlers of the query engine (e.g., query start/completion handlers). We send query logs to PlazmaDB through Fluentd, which resides in the machine as a  daemon process. We use Fluency \cite{fluency}, another open-source product by our engineer, for efficiently sending logs from JVM programs to Fluentd. 

As we mentioned earlier, OSS query engines are frequently updated, and we also need to apply changes related to our own extensions. Isolating the query engine code from the underlying storage service is important because the APIs of storage services are less frequently updated. For example, the core API of S3 has not been changed for a long time since 2006. Another benefit of isolating the storage service is that we can operate multiple query engines, including Trino, Hive, Spark \cite{10.1145/2934664}, etc., on top of the same PlazmaDB. This capability is especially useful for query simulation, which needs to run multiple versions of the query engine using control (current version) and test (next version) clusters attached to the same production storage. Blue-green deployment \cite{bluegreen} also becomes possible to safely migrate the customer traffic from the current cluster (blue) to a new cluster (green) if the same storage is shared between clusters.


\subsection{Secure Query Simulator}

For simulating customer queries, we must ensure the security of the customer data and protect the privacy of the customer query results.
As our employees are not allowed to see any customer data, we needed a secure environment that protects the customer data when engineers are debugging customer queries. We also need to protect our employees from accidentally looking at or modifying customer data. To run query simulation, we are issuing short-term special tokens, which are effective only within our private network and provide read-only access to the data and a scoped write-permission to a temporary storage inside the customer storage. This ensures the data used for testing update queries will never leak outside the customer storage boundary. 

To protect the privacy of our customer query results, the query simulator rewrites queries to produce only the checksum of the query results. The simulator also rewrites update queries to use only the temporary storage. To compare the query results whose record order is undefined, we need to use order-insensitive checksum, for example, by taking XOR of record values. As we need to compute the checksum of thousands of simulated queries per customer, we use order-insensitive \textsf{checksum()} function in Trino to perform the query simulation and checksum computation simultaneously. This approach works efficiently because it removes the overhead of writing the query results to a temporary storage.





\begin{table}[t]
\centering
\scriptsize
\begin{tabular}{ |c|l| }
 \hline
 \rowcolor{lightgray} 
 \textbf{Query signature} & \textbf{Corresponding SQL statements}\\
 \hline
 S(T) & SELECT ... FROM ...  \\ 
 S[*](T) & SELECT * FROM ... (select all columns) \\
 G(S(T)) & SELECT ... FROM ... GROUP BY  \\  
 S(LJ(T, T)) & SELECT ... FROM .. LEFT JOIN ... \\
 WS[A(a,S(T))] & WITH a AS SELECT .. (define aliases to queries) \\
 O(S(T)) & SELECT ... ORDER BY \\
 CT(S(T)) & CREATE TABLE AS SELECT ... \\
 I(S(T)) & INSERT INTO ... SELECT ...\\
 E(S(T)) & SELECT distinct ... FROM ... (duplicate elimination) \\
 U(S(T),S(T)) & SELECT ... UNION ALL SELECT ...\\
 \hline
\end{tabular}
\caption{Examples of query signatures}
\label{tab:querysig}
\end{table}

\section{Analyzing Query Logs}
\label{sec:analyzing}

Our query logs stored in PlazmaDB contain raw query statements, query IDs, execution start and end time, query duration, query plans, resource usage stats such as CPU, memory, partition read/write, etc. As we have millions of query logs per day, analyzing these query logs also needs distributed query processing using columnar storage. In this section, we will introduce query signatures for clustering queries, which can be used for monitoring the query performance and building customer-specific SQL benchmarks.

\subsection{Query Signatures}

To create clusters of SQL queries, we need to abstract away subtle differences of query expressions. First, we extract the logical plan of a SQL statement, then replace the plan tree nodes with simple letters, such as, select (\textsf{S}), group by (\textsf{G}), joins (\textsf{J}), input table (\textsf{T}) etc. For describing nested SQL statements, query signatures enclose sub queries within parentheses. For example, the query signature of a SQL, \textsf{SELECT c FROM A}, becomes \textsf{S(T)}. Table \ref{tab:querysig} shows more examples of query signatures. We implemented \textsf{TD\_QUERY\_SIG(query)} UDF for computing the query signature of a given SQL statement. With this UDF, millions of query signatures can be computed efficiently with Trino.

Query signatures can also have the input and output table names of queries as suffixes so that we can create clusters of queries accessing the same table sets. For example, a SQL, \textsf{INSERT INTO A SELECT .. FROM B}, which reads data from table B and writes the results to table A has a query signature \textsf{I(S(T)) B->A}. 
Some customers manage intermediate table names with date-time patterns, e.g., table-2022-02-01, table-2022-02-02. To create clusters of such queries that are almost identical between different days, we apply masks for date-time patterns. For example, we can aggregate table-2022-02-?? signatures as a single pattern, table-X.


\subsection{Building Customer-Specific Benchmarks}

We build customer-specific benchmarks by taking a representative query (e.g., the latest one) from each query cluster grouped by query signatures, which can be computed at ease with a Trino SQL using \textsf{TD\_QUERY\_SIG(query)} and aggregation functions against our query logs. 
There is always a trade-off between the test coverage and test efficiency. If we run more queries to improve the coverage, we need to wait for a long time until the simulation completes and the cost for running queries increases. Our query simulator accepts various tuning parameters for controlling the test size, such as input query history ranges, target customers and queries to test, etc. If we need more precise query patterns and if the customer is using our hosted workflow engine Digdag \cite{digdag}, we can use its workflow task IDs or query template IDs for grouping recurrent queries. If customers manage workflows with their own job management system or scripts, we don’t have any useful IDs for grouping queries. Clustering with query signature works as an unsupervised method even if there are no such identifiers. 

Note that, query signatures are just one of the features for clustering similar queries. We can also employ other metrics, such as detailed query expression values, resource usage, running time, etc. Shaleen et al. \cite{10.14778/3430915.3430931} studied a query workload compression technique based on feature sets chosen by experts. Finding such feature sets to balance the workload size and the representativity of the customer queries is our future work. In our experiences, debugging bugs in query engines is more challenging than improving the test coverage. As bugs often happen in unexpected query patterns, quickly finding such queries is a key to address potential regressions of the query engine service.

\subsection{Implicit SLOs of the Query Performance}




Query signatures can also be used for providing a measure of performance consistency. As both of us and customers don't have any explicit definitions about the stability of the query performance, we defined implicit SLOs (service-level objectives) for individual query groups created by query signatures. We compute the median (p50) of query running times in each query group\footnote{Since the average running time can be disturbed by extreme values, using median provides more robust measures.} and median-absolute deviation (MAD) \cite{199313}, then define an SLO range of a query running time as (median) $\pm$ 3MAD, indicating that 99.7\% of queries finishes within this performance range assuming the normal distribution. If the query running time exceeds this range, we call it an SLO violation. Practically, we can use \textsf{approx\_percentile($x$, 0.50)} of Trino, which can efficiently compute an approximate median value without sorting the entire input values. CoV (Coefficient of Variation) = MAD / median is useful to see the variance of the query performance within a cluster. If CoV > 1, the query performance within the cluster is unstable, so estimating SLOs from its historical stats is not preferred. In this case, we need to add more features to refine the quality of the query cluster.

\begin{figure}[t]
\includegraphics[width=0.43\textwidth,keepaspectratio]{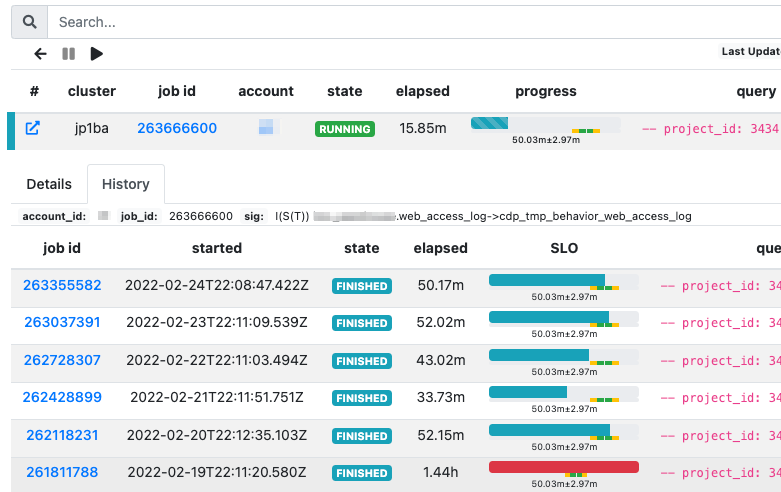}
\caption{Monitoring query performance with implicit SLOs}
\label{fig:slo}
\end{figure}

Figure \ref{fig:slo} shows a screenshot of our query progress monitor that displays the SLOs of individual queries. The blue bars indicate that queries are running within the SLO ranges, and the red bars show that the queries are slower than usual and violate implicit SLOs. 
Green and yellow lines at the bottom of the progress bars show the median $\pm$ 1MAD, $\pm$ 2MAD ranges, respectively. This monitor can show the historical stats of the queries that have the same query signature. This historical monitor has improved our experience of investigating query performance issues as we can see the user's implicit expectation for the query performance.

\section{Migrating Production Queries}
\label{sec:migrating}

By using the query simulator, we have successfully migrated production customer queries for years. Upgrading query engine versions containing incompatible changes is always a big effort as it requires communication with customers, but other internal changes also become possible with the simulator, such as upgrading JDK version from Java 8 to Java 11, Linux OS upgrade, CPU type change from Intel to cost-efficient Arm Graviton2 processors \cite{graviton2}, etc. 
We also found several bugs in Trino OSS with the query simulator 
and reported them to the OSS community (Table \ref{tab:bugs_found_by_simulator}). The most of these issues have already been fixed, but some of them are specific to query patterns or input-data types. Running customer-specific benchmarks using real data sets has been helpful to find such corner-case bugs.

\begin{table}[t]
\scriptsize
    \centering
    \begin{tabular}{|c|l|}
    \hline
    \rowcolor{lightgray}     
    \textbf{GitHub Issue} & \textbf{Description} \\
    \hline
    \textsf{\href{https://github.com/trinodb/trino/issues/8027}{\#8027}} &
    'Multiple entries with same key' error on duplicated grouping of literal values\\    \textsf{\href{https://github.com/trinodb/trino/pull/10764}{\#10764}} & 
    Missing shallowEquals() implementation for SampledRelation\\  
    \textsf{\href{https://github.com/trinodb/trino/issues/10861}{\#10861}} &
    Query fails if IS NOT NULL is used for information\_schema \\      \textsf{\href{https://github.com/trinodb/trino/issues/10938}{\#10938}} &
    Predicate push down doesn't work outside the scope of sub query \\
    \textsf{\href{https://github.com/trinodb/trino/pull/11259}{\#11259}} &
    TRY should handle invalid value error in cast VACHAR as TIMESTAMP \\
    \textsf{\href{https://github.com/trinodb/trino/pull/12199}{\#12199}} & IllegalStateException in ScopeAware.scopeAwareComparison()\\
    \hline
    \end{tabular}
    \caption{A list of Trino OSS bugs found by query simulation}
    \label{tab:bugs_found_by_simulator}
\end{table}

\subsubsection*{Running Query Simulations}

Even though we have reduced the number of queries to test with query signatures, we still need to simulate from 100k to 1M query patterns out of 45 million queries running in a month. Unless there is a major version change (e.g., upgrading Trino OSS version, optimizer changes, etc.), we usually use 7-day range of the query history to build a customer-specific benchmark to accelerate the query simulation. When we need to report full simulation coverage, we use the entire query history of the last month (about 30 days) to cover monthly recurrent queries. If we need to upgrade only for a small set of customers, we can run a targeted query simulation for them.

\subsubsection*{Evaluating Query Simulation Report}

\begin{figure}[t]
    \centering
    \includegraphics[width=.45\textwidth]{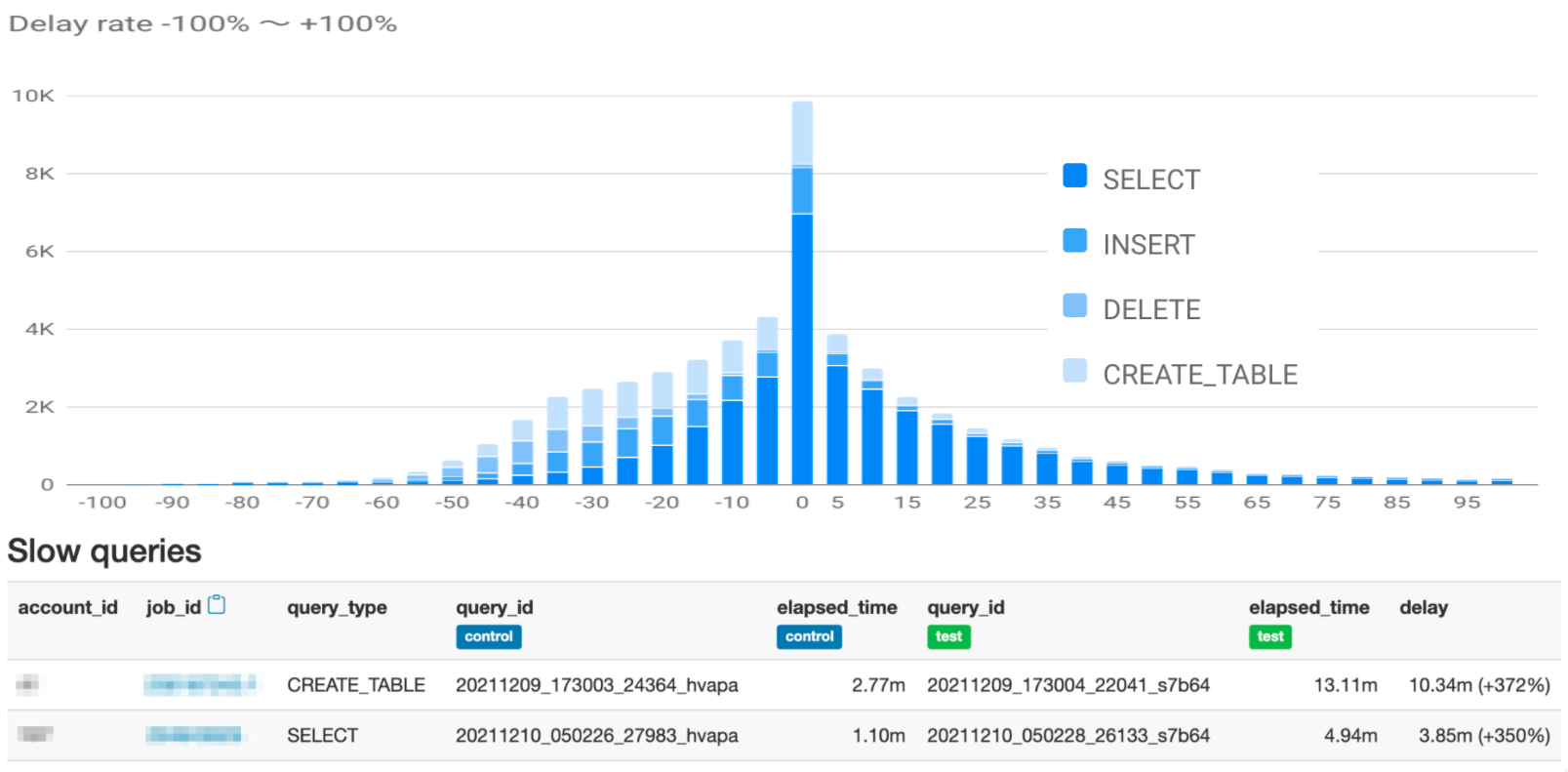}
    \caption{Query simulation report for comparing the performance differences of control and test clusters}
    \label{fig:simulation_report}
\end{figure}

Figure \ref{fig:simulation_report} shows an example of a query simulation report that compares the query performance between control and test clusters launched in a pre-production environment. This report also shows a list of queries that are significantly slower. After running the query simulation, we usually start an investigation from these queries to identify the cause of the performance regression. If we see a big difference in the number of accessed partitions, we can suspect a bug in recently changed query optimization rules. In order to report such optimizer issues to the open-source community without revealing the customer data and queries, we have developed Trino compatibility checker \cite{trino-comp}, which runs the same query in multiple versions of Trino and detects a version that introduced the regression. Even if we can't share the exact query and data, reporting a specific version and reproducing the same error with similar queries have been working well for getting help from the community.

The query simulator also produces a query correctness report by comparing the checksum of the query results. Depending on the types of queries, queries may produce non-deterministic result, especially if the query has rounding of floating value computation or non-deterministic operators. An example of non-deterministic operators is \textsf{max\_by}, which returns an arbitrary record value among the ties, so the result may change depending on the record processing order. UDFs looking up external services may also dynamically change the returned values. Other such expressions include \textsf{now()}, \textsf{random()}, window functions, approximate counting operators which use randomized algorithms. As query simulation reports have a lot of such non-deterministic results, properly labeling the presence of non-deterministic expressions is helpful to reduce the burden of engineers evaluating the simulation report.

\subsubsection*{Scheduling Releases}

Releases around holiday seasons are not generally preferred
not only because people are off, but also these seasons usually have the highest volumes of data in the year related to shopping. If some of their data pipelines is broken, the companies will lose a lot of revenue. Note that holiday seasons are different by countries; In US, Independence Day (July 4th), Thanksgiving (November 24). In Japan, Golden week (weeks around May 3 - May 5), Obon (middle August), new year holiday (the first week of a new year), etc. Releases that have no user-facing changes are acceptable, such as adding new functionalities or upgrading internal library versions, as long as the quality of the releases is ensured.

To control the blast radius of new releases, we practice canary deployment, which gradually migrates the customer traffic to a new query engine version for each user group and deployment region. If we find any problem during the canary deployment, we can quickly switch back the traffic to the previous version of the service. When customers need to take actions for fixing their queries due to the incompatibility between versions, we may spread the release schedules between individual customer groups to allocate more grace period.

\section{Discussion \& Future Work}
\label{sec:discussion}

\subsubsection*{No Look Debug of Queries} 
Identifying the cause of inconsistent query results is still challenging because we can see only checksum of the query results and query plans, but we can't directly see the customer data and query results. To locate where the error is happening, we need to breakdown queries into tiny pieces. In case there is a regression in query optimization rules, comparing query plans between versions is also helpful.
To find target sub queries for the investigation, we often try to locate suspicious operators or expressions in the query. We need more automation for helping such drill-down investigation.


\subsubsection*{Testing Query Rewrite Optimization} Maintaining the query optimization rules has been the most challenging problem. As query plans can have hundreds of stages, debugging optimization issues relying on human eyes has its own limitation. Trino uses a visitor pattern \cite{10.5555/186897} in Java for rewriting logical and distributed plan tress.
Propagating optimization contexts within the visitor pattern leads to complex code and we often see non-covered tree patterns during tree rewrites. And also, if a new type of plan tree nodes is introduced, we need to review the coverage of query pattern traversals. This type of errors often produces non-optimized plans. To detect such issues during query simulation, we added steps for comparing metrics around the number of scanned partitions, memory usage, performance distributions etc. 

\subsubsection*{Long Running Queries} We have learned that having some query running time limit eventually helps the users and improves maintainability of the service. It’s not uncommon for users to process large volumes of data in a single query that may take several hours. Such long-running queries, however, are hard to maintain for the users and become slower as the data size grows. For long running queries, even if we notice any SLO violation after several hours, it's already too late to recover. 
In addition, while Hive is suited to long-running queries as it supports recovery from failed tasks, Trino uses a fixed set of worker nodes for processing queries, so queries fail if one of the worker nodes crashes. Even though our system supports retrying failed queries from scratch, the retry cost is expensive for long-running queries.
Initially, we provided the query service without any limitation for the query running time, but gradually introduced 2 day running limit, and in 2022, the limitation became 6 hours for Trino. We are planning to reduce the max query running time of Trino to 2 hours so that the customers can manage their queries into small pieces and reduce the turn-around time for debugging queries.

\subsubsection*{Query Simulation Isn't Free} Running query simulations using production data set is not free in terms of its running cost as well as pressures to other production services. Query engine services usually have other dependent APIs for user authentication, table catalogs, storage (e.g., PlazmaDB, S3), etc. If we run a lot of query simulations concurrently, it will add pressures to these dependent APIs,
which may trigger incidents at production, so care should be taken to manage the additional API requests coming from the query simulation traffic. To minimize such pressures, we narrow down the input data volume size of test queries by restricting the target time ranges while considering seasonal effects, such as monthly batch queries, etc. We also need more improvement to the query simulator to avoid running hog queries and redundant queries which will not improve the test coverage. Building synthetic data sets that are disconnected from the production data \cite{emam2020practical} is an option to isolate the simulation traffic from the production system, but we need further study to properly capture input-data specific issues in query engines while following various regulations under privacy laws (e.g., GDPR, CCPA, HIPPA, etc.) when building synthetic data generated from production data set models.

\section{Conclusions}
\label{sec:conclusions}

We presented the architecture of our query engine service on the cloud. As the number of customers relying on the correctness and stability of the query engines increases, upgrading the query engine versions becomes challenging. To maintain the quality of the query engine service, we need to continuously collect query logs whose schema is evolving over time, and build a secure environment for running customer-specific benchmarks using the real production data sets. 
We believe that our system architecture and the best practices described in this paper are useful resources for building your own data platform on the cloud.

\bibliographystyle{ACM-Reference-Format}
\bibliography{dbtest2022}


\end{document}